# Fractal flow patterns in hydrophobic microfluidic pore networks: experimental modeling of two-phase flow in porous electrodes

*Viatcheslav Berejnov, Aimy Bazylak[1], David Sinton, and Ned Djilali*

*Department of Mechanical Engineering and Institute for Integrated Energy Systems, University of Victoria, Victoria, V8W 3P6, Canada*

[1]*Department of Mechanical & Industrial Engineering Faculty of Applied Science & Engineering, University of Toronto, 5 King's College Road, Toronto, ON Canada M5S 3G8*

Correspondent author: berejnov@gmail.com

**Abstract**

Experimental two-phase invasion percolation flow patterns were observed in hydrophobic micro-porous networks designed to model fuel cell specific porous media. In order to mimic the operational conditions encountered in the porous electrodes of polymer electrolyte membrane fuel cells (PEMFCs), micro-porous networks were fabricated with corresponding microchannel size distributions. The inlet channels were invaded homogeneously with flow rates corresponding to fuel cell current densities of 1.0 to 0.1 A/cm$^2$ (Ca $10^{-7} – 10^{-8}$). A variety of fractal breakthrough patterns were observed and analyzed to quantify flooding density and geometrical diversity in terms of the total saturation, $S_t$, local saturations, $s$, and fractal dimension, $D$. It was found that $S_t$ increases monotonically during the invasion process until the breakthrough point is reached, and $s$ profiles indicate the dynamic distribution of the liquid phase during the process. Fractal analysis confirmed that the experiments fall within the flow regime of invasion percolation with trapping. In this work, we propose to correlate the fractal dimension, $D$, to the total saturation and use this map as a parameter for modeling liquid water transport in the GDL.

## 1. Introduction

A thorough understanding of multiphase flow in the gas diffusion layer (GDL) is critical for improving the design and optimization of polymer electrolyte membrane fuel cells (PEMFCs). The GDL, also known as the gas diffusion medium or porous transport layer, serves multiple mutually dependent functions that are essential for PEMFC operation. The GDL enhances the removal of excess water from the catalyst layer, mechanically supports the membrane electrode assembly, and provides pathways for both gaseous fuel and electron transport. This multifunctionality makes the optimization of GDL properties highly challenging, yet equally sought after by PEMFC manufactures. Unfortunately, the opaque nature of this material (Figure 1) greatly limits the accessibility for direct optical observations of the through-plane liquid water transport. Thus, the nature of multiphase flow in the GDL still remains a topic of major curiosity. The lack of predictive multiphase transport models for the GDL inhibits the feedback between GDL design and water management optimization techniques needed for improving PEMFC performance.

Due to the lack of realistic two-phase flow data, the multiphase PEMFC continuum models have to date relied primarily on empirical measurements of water transport in unconsolidated sand (1), though characterization studies conducted in PEMFC specific porous media have appeared recently (2, 3). In contrast, in this work we directly employ transparent pore networks with wettability properties similar to the GDL to provide insights on liquid water transport in the GDL during realistic PEMFC operating conditions. We demonstrate that combining saturation and fractal dimension data on one diagram yields a correlation between the bulk (saturation) and structural (fractal dimension) transport characteristics in the GDL.

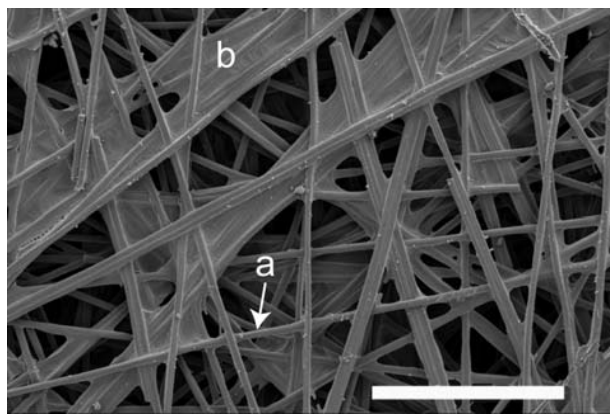

**Figure 1:** An SEM image of Toray carbon paper (TGP-H-060 10wt. %). The diameter of a carbon fiber (a) is ~8 μm. (b) Local patches of PTFE-coated binder. The length bar is 200 μm.

This paper is organized as follows. The pore network approach is introduced, followed by a discussion of microfluidic pore network simulations of the GDL. The methods of fabricating the pore network microfluidic chips are described. Next, we discuss and analyze the observed flow patterns. The liquid water transport observations that mimic *in situ* GDL conditions are presented and compared with our theoretical pore network results.

## 1.1 Pore network approach

Substituting a complex porous media with a network of inter-connected pores and throats (4) is a powerful numerical method for modeling transport in porous materials (5, 6). In 1977, this idea was first realized experimentally using a lithography technique (7, 8) for fabricating microchannels in a polymer resin. Throughout the following decades other methods of fabricating pore networks were proposed and tested: glass etched networks (9), glass micro-machined networks (10), networks of packed glass spheres (11), and 3D layered glass etched networks (12). Despite the fact that these techniques produced transparent pore networks, the topology, shape, and distribution of pore channels were not suitable for GDL modeling. Furthermore, these networks were designed to obtain high chemical affinity to either injected or ejected fluid, which resulted in the non-wetting conditions for the opposite fluid and low or zero wettability hysteresis for the triple contact line. In contrast, the GDL is a hydrophobic media that exhibits *high* wettability hysteresis, which dramatically affects the capillary flow behaviour.

In earlier literature (13, 14), it was found that depending on flow conditions, the frontal and internal structures of invaded or drained fluid patterns in pore networks (fabricated from resin and glass materials) were either continuous or fractal. Mathematical models describing these processes were developed, and direct numerical simulations agreed well with experiments (13). It was also shown that fluid substitutions in a particular pore network resulted in a variety of flow patterns (13). Generally, these patterns can be organized in three classes depending on the model of how liquid fills the pore space. Invasion due to *viscous fingering* occurs when the invasion is locally non-correlated and the paths are non-overlapping; the diffusion limited aggregation model (DLA) describes this invading scenario (15). Invasion due to *capillary fingering* occurs when invading paths are still non-correlated, but experience overlap resulting in secondary phase entrapment. This behaviour corresponds to the invasion percolation with trapping model (IPT) (16). The third class is characterized by a continuous invasion pattern, when the invading paths are locally highly correlated. Two dimensionless numbers govern the structure of flow patterns: capillary number $Ca=\eta_2 U/\sigma$, and viscosity ratio $M=\eta_2/\eta_1$, where $\eta_1$, $\eta_2$, $U$, and $\sigma$ are the dynamic viscosities of fluids 1 (ejected) and 2 (injected), mean flow velocity, and interfacial tension, respectively. An extensive amount of data, analysis, and theories explaining the different regimes of invasion, two-phase patterns, and structures may be found in the following reviews (17, 18). A number of studies have recently adopted with some success the numerical pore network methodology to investigate two-phase flows relevant to PEMFC materials and operating conditions (19-22).

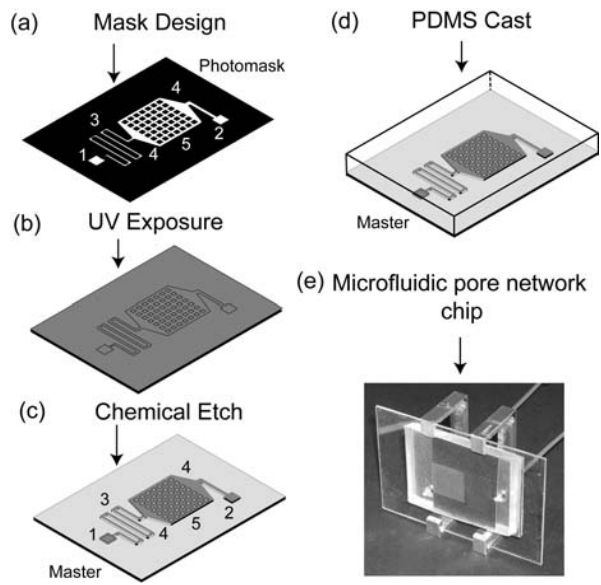

**Figure 2:** Fabrication procedure for microfluidic pore network chips. (a) A high resolution pore network photomask is produced. (b) A flat substrate coated with a thin photoresist layer (~ 25μm) is exposed to UV light. (c) Unexposed photoresist is chemically etched away leaving a planar 3D pore network structure (master). (d) A PDMS elastomeric replica of the pore network pattern is cast and assembled into a microfluidic pore network chip (e).

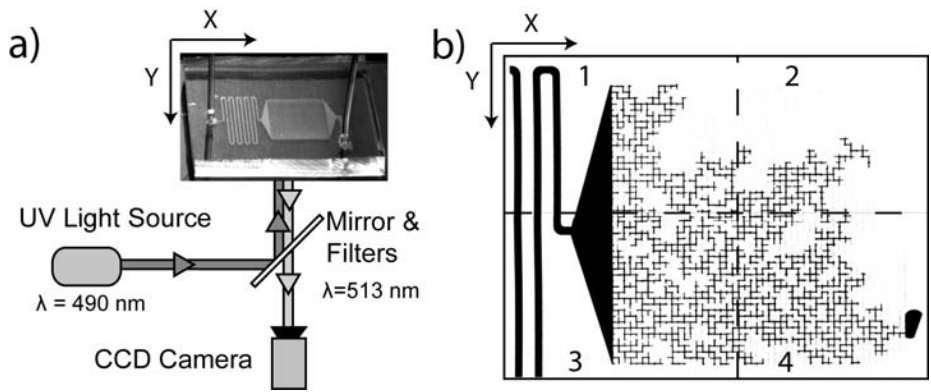

**Figure 3:** Schematic of the experimental apparatus: (a) A fluorescence enabled inverted microscope is employed with an automated sample stage for programmable X-Y positioning. (b) The complete flow pattern is captured through a collection of four images captured in rapid sequence. Each compiled flow pattern is assigned an image index, $n$, within the stack of $n_{max}$ flow pattern images.

**1.2 Employing microfluidic pore networks to model the GDL**

To date there are three main non-invasive techniques for observing liquid water accumulation in operating PEMFCs: Nuclear Magnetic Resonance (NMR) imaging, (23, 24), neutron imaging (25), and X-ray imaging (26). Please refer to (Bazylak, 2009) for a thorough review. However, these techniques involve very expensive equipment and time consuming data acquisition methods, which are not feasible for many fuel cell laboratories. Most importantly, the high lateral resolution needed to resolve transport in the GDL with pores ranging from 10-100 μm has yet to be achieved.

The structural and dynamic insights of microscopic flow patterns at the pore-scale level are crucial for developing a predictive two-phase mathematical model of the GDL. Recently, for example, two different scenarios of breakthrough flow patterns in the GDL were proposed. In the *converging capillary tree* framework proposed by Nam and Kaviany (27), liquid water initially forms micro-branches in the GDL, which further spontaneously converges into a single main percolating root. Litster et al. (28) proposed the *fingering and channeling* framework whereby micro-branches of liquid water grow and percolate within the GDL independently, forming structures resembling isolated fingers. Since these scenarios cannot be verified through direct optical visualizations, the transparent porous medium employed in this work has the potential to provide further insight into quantitative liquid water transport behaviour in the GDL.

In order for a transparent microfluidic pore network to accurately model the GDL, several qualifications must be met: a) the pore network has a tunable pore and throat topology, b) the method of flow visualization is non-invasive and employs white light, c) the method of flow classification allows for a numerical distinction between different flow patterns, d) the flow patterns are deterministic, e) the flow rates through the pore network should be representative of *in situ* GDL conditions, and f) the network structure can be verified with numerical pore network simulations. An approach that can satisfy many of these requirements was recently demonstrated by our group (29). The approach involves utilizing microfluidic patterning techniques to generate micromodels for the study transport in porous media. In that proof of concept work, applicability of the method to study both oil-air and water-air imbibition processes was established.

In this paper, a microfluidic network approach is applied to the study of the invasion of liquid water in hydrophobic porous media. The characteristics of the microfluidic network are matched as closely as possible to that of the GDL in PEMFCs. Employing fluorescence imaging and image analysis, the dynamics of the injection process are studied as a function of flow rate. Results are analyzed using total saturation, local saturation profiles (and their evolution over time), and fractal dimension, and compared with previous works in the areas of GDL transport and invasion percolation theory.

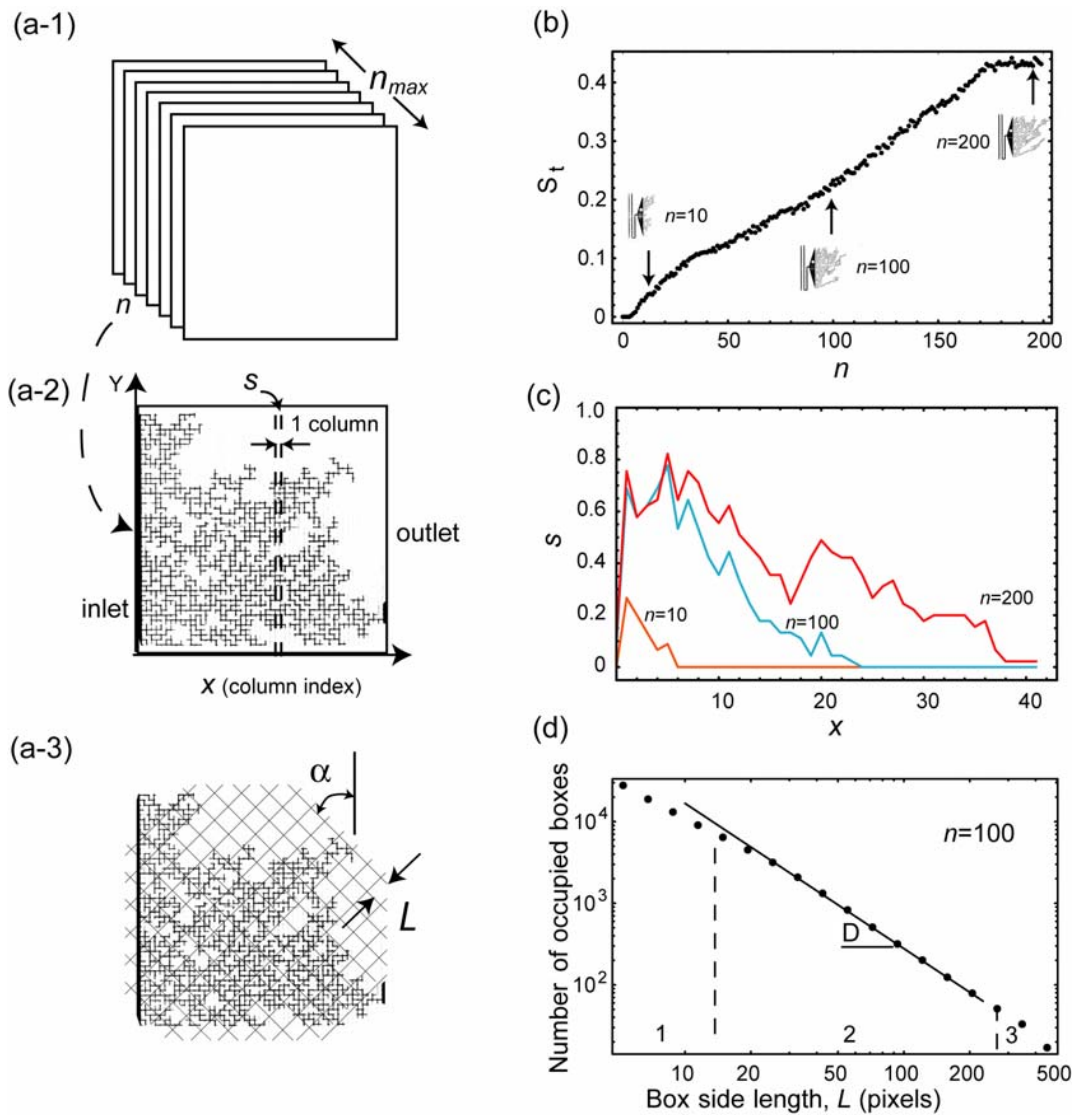

**Figure 4:** Image analysis procedure with sample data: (a-1) A stack of $n_{max}$ pore network images obtained during the filling process. (a-2) A sample image showing column-wise averaging of the local saturation, $s$. The total saturation, $S_t$, is calculated from the complete network pattern. (a-3) Fractal dimension analysis ($D$). (b) $S_t$ as a function of image index, $n$. (c) $s$ profiles for selected image indices corresponding to flow pattern inserts shown in (b). (d) Calculation of $D$ for one image ($n = 100$).

## 2. Experimental
### 2.1 Fabricating microfluidic pore networks

The microfluidic networks employed in this work were fabricated with a conventional soft-lithography technique (30, 31). In an earlier work, we developed some modifications to the above technique in order to build large pore networks with variable material surface properties (Figure 2) (29). The fabrication of the microfluidic pore network consisted of four general stages: designing and printing a pore network photomask (Figure 2 (a)), fabricating a master (Figure 2 (b, c)), fabricating a PDMS replica (Figure 2 (d)), and assembling the parts into a microfluidic pore network chip (Figure 2 (e)). The final device was a transparent elastomeric microfluidic chip, which consisted of a system of rectangular (25 x $L$ x $W$ μm) microscale channels embossed in a PDMS polymer (Figure 2 (e)). The length, $L$, and width, $W$, of the channels were distributed in the following ranges: $L \in (200\text{-}300 \mu m)$ $W \in (10\text{-}130 \mu m)$, respectively. The networks were rectangular with overall dimensions of approximately 15x15 mm and contained approximately 5000 channels.

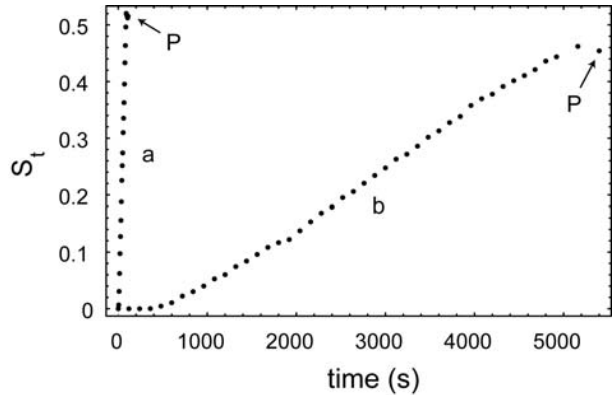

**Figure 5:** The total saturation of a pore network with respect to filling time for two different flow rates: (a) $Q = 0.5$ μL/min and (b) $Q = 0.01$ μL/min. $P$ denotes the breakthrough moment.

### 2.1.1 Designing, integrating, and printing pore-network photomasks

We fabricated a large number of representative microfluidic pore network patterns with a high pore/throat density that regularly deviated from one another. The development of the network pattern was divided into two stages that dramatically simplified the mask design. The first stage consisted of designing the mask manifold using a drawing software program (AutoCAD). As shown in Figure 2 (a), the mask manifold consisted of: delivery channels (1, 2), pressure-drop channels (3), inlet/outlet manifolds (4), and a zone for the pore network in the center (5). During the second stage, pore networks were designed with an in-house numerical software package (32). By varying the pore network pattern design, the topology of the network was easily modified: pore-to-throat volume ratio, pore connectivity, throat distribution, and even superstructure (32).

Once the pore network patterns were designed, generated and tested, they were inserted into the AutoCAD template. AutoCAD drawings were printed on transparencies with a high resolution (5K, 10K, and 20K dpi) commercial printing service (Outputcity, California). These transparencies were employed as photomasks for fabricating the negative masters.

Before embedding the coordinates of a pore network into an AutoCAD file, liquid water invasion was simulated with an invasion percolation with trapping algorithm (33) to identify the expected deterministic flow patterns. Further details on this pore network model can be found in (19).

### 2.1.2 Fabricating masters and PDMS replicas

The negative masters (Figure 2 (c)) of microfluidic pore networks were fabricated using the soft lithographic technique described in earlier work (29, 30, 34) with process-specific information provided in (29). The structural properties of the pore network depended on the chip fabrication procedure. The thickness, $h$, of the network pattern on the master (Figure 2 (c)) scaled the size of the fabricated channels. Varying $h$ altered the hydraulic radius $R_h = A/d$ of the channels, where $A$ and $d$ were the cross-sectional area and characteristic diameter of the channel, respectively.

A PDMS polymer (Sylgard 184, Dow Corning, NY) was cast against the master as shown in Figure 2 (d). After the PDMS was cured, it was cut with a blade and peeled away from the master to yield a replica containing a positive structure of the network channels. The inlet and outlet ports were punched with rounded holes for holding plastic capillary tubing (1/16[th] inch OD Teflon tubing, Fisher Scientific). The thickness of all replicas was approximately 5 mm, which enabled firm connections to chip ports without additional flow adapters.

The PDMS replica was combined with a PDMS covered glass slide, as shown in Figure 2 (e). A thin layer of PDMS on the glass slide provided a flat and thin substrate for our pore networks. The elastic replica and the substrate were compressed with clamps to provide a firm contact for leakage prevention. With this reversible seal, the chip was opened, cleaned, and treated for multiple uses.

### 2.2 Recording flow patterns

All pore networks were initially dry, and aqueous fluorescein solutions (0.25 mM of fluorescein dye from Invitrogen, Canada) were injected into the pore networks during the drainage process, while air was injected to observe the invasion process (See Table 1 for physical properties). Low liquid flow rates ($Q = 0.01\text{-}2\mu L/min$) were chosen to mimic *in situ* PEMFC GDL conditions. Liquid injection was controlled with a programmable syringe pump (70-2211 Harvard Apparatus, USA). For liquid injection rates of 0.5-2µL/min and 0.01-0.5µL/min, 1mL plastic syringes (BD, USA) and 100µL glass syringes (Hamilton, USA) were employed.

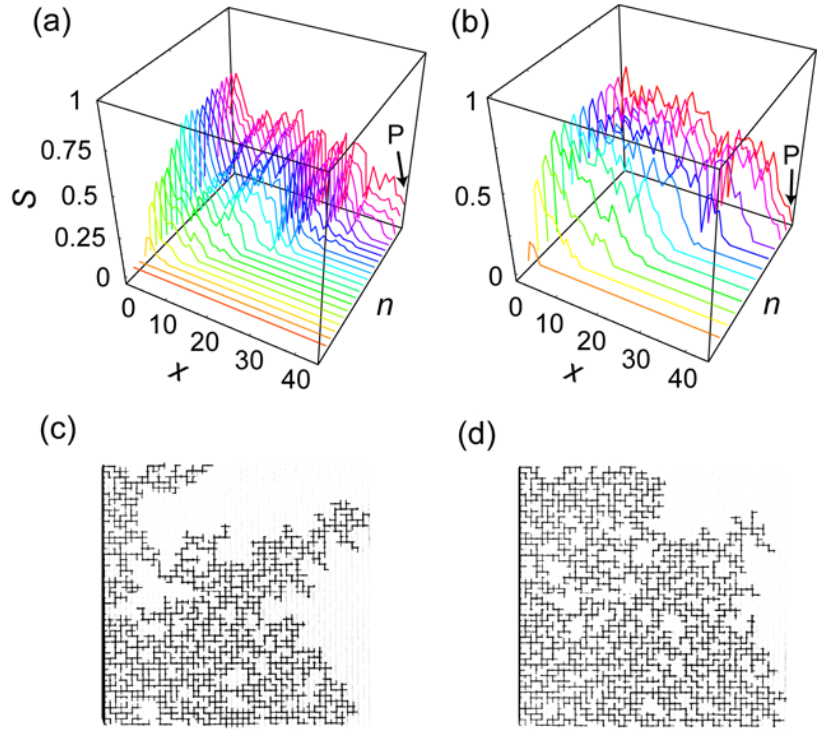

**Figure 6:** Evolution of the local saturation with respect to the image index, *n*, and column index, *x*, for two flow rates: (a) $Q = 0.01$ µL/min and (b) $Q = 0.5$ µL/min. *P* denotes the breakthrough moment. Images (c) and (d) are the breakthrough patterns at *P* shown in (a) and (b), respectively.

Fluorescent microscopy provides a high signal-to-noise ratio in flow images. We collected flow patterns using a fluorescence inverted microscope (Leica DMI 6000B, Germany) and a 1.25x0.04 objective (HCX PL Fluotar, Leica, Germany). The field of view was approximately 0.8 cm, and a set of four images captured in rapid sequence provided a complete network image, as illustrated in Figure 3. The maximum achievable image resolution was X: 2688 pixels and Y: 2047 pixels, which corresponded to a physical length of 22 mm x 16.76 mm, respectively. The number of pixels per throat width varied from 4 to 12 pixels depending on the pore network design.

We observed reproducible flow patterns with the same microfluidic network. This repeatability indicated that initial flow disturbances and small assembly perturbations had an insignificant impact on the final flow pattern. Between the repeated trials with the same microfluidic network, the chip was disassembled, rinsed, dried and reassembled. The series of flow patterns were registered and compared, and it was found that the flow patterns were remarkably similar. The degree of flow pattern reproducibility was quantified by the difference between the number of filled throats for two different experimental realizations. We obtained approximately 96-99% reproducibility. This reproducibility validated the deterministic nature of the theoretical invasion percolation with trapping model that we employed for this study (19).

## 3. Methods
### 3.1 Characterizing flow images

We analyzed a sequence of $n_{max}$ images by calculating for each complete network image: the total network saturation, $S_t$, the local saturation, $s(x)$, and the fractal dimension, $D$, as shown in Figure 4. The total network saturation was defined as $S_t = N_{inv}/N_{tot}$, where $N_{inv}$ and $N_{tot}$ were the number of invaded throats and the total number of throats for each given image (Figure 4 (a-2) and (b)). The local saturation was defined as $s(x)=N_{inv}(x)/N_{tot}(x)$, where $x$ was the column index, running between 1 and 49 from the inlet to outlet, and $N_{inv}(x)$ and $N_{tot}(x)$ were the number of invaded throats and total number of throats for a given throat, $x$, respectively (Figure 4 (a-2) and (c)).

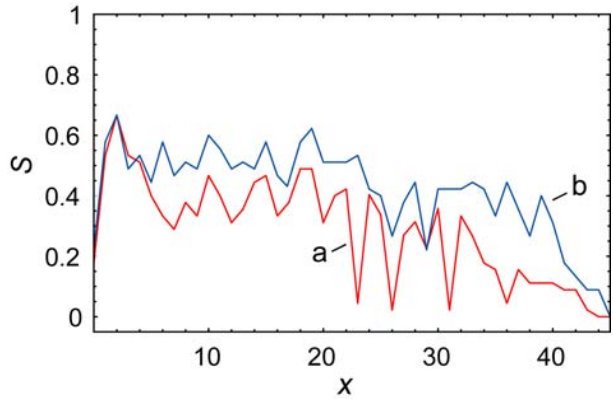

**Figure 7**
The saturation profiles taken from Figure 6, which correspond to the breakthrough moments for two pump flow rates: (a) 0.01μL/min, where Ca ~ $10^{-8}$ and (b) 0.5μL/min, where Ca ~ $10^{-6}$.

We employed the fractal dimension, $D$, to quantify the structural differences between flow images. In simplified terms, the fractal dimension is a measure of how the invading phase 'takes up space' within the microfluidic pore network (18). Mathematically, the fractal dimension was defined as $D =1-\log(F)/\log(L)$, where $F$ was the total extent (or length) of the considered pattern, and $L$ was the metric (or length scale) on which it was measured (18). Two limiting cases illustrate the conception of the fractal dimension: a line or circle has a fractal dimension of $D = 1$. A solid area or disk has a fractal dimension of $D = 2$. For a given flow pattern within our microfluidic network, the fractal dimension is roughly expected to lie in the range $(1 < D < 2)$ and may be calculated from experimentally obtained fluorescence images (29).

To calculate $D$, we applied a box counting method (Figure 4 (a-3) and (d)), using both our in-house numerical software package (LabView) and a commercial software package (Benoit 1.3). In the box counting algorithm, a minimum number, $N$, of boxes of length, $L$, is determined necessary to cover the fluorescent pattern on the image. If the fluorescent structure is fractal, then its dimension will be in the range $D \in (1 < D < D_n)$, where $D_n$ is the fractal dimension of the entirely filled network (~ 1.9 for all networks). The fractal dimension, $D$, may be determined from the relation $N \sim (1/L)^D$ (18), which was obtained from a plot of the log-log scale (Figure 4 (d)). We applied the box counting method at different angles of our images and averaged these results to increase the accuracy of our $D$ measurements. Our fractals had two natural length scale limits, where the largest scale represents the network size and the smallest scale the throat period. We considered only the middle length scale region (2) in Figure 4 (d) when calculating $D$.

## 4. Results
### 4.1 Effect of flow rate on the total saturation, $S_t$

The total saturation, $S_t$, was a convenient parameter for characterizing the average for the entire pore network and was employed in the lumped approach for calculating two-phase transport in the GDL. $S_t$ was defined as the ratio of filled throats to initially void throats. Figure 5 shows the total saturation versus an injection time for two different injection rates, $a$ and $b$. Every point on this figure was derived from an image consisting of a flow pattern similar to that presented in Figure 4 (a-2). The saturation $S_t$ was calculated by summing the number of completely filled and 2/3-filled channels and dividing this sum by the number of initially void channels in the network (~4900). The total void volume of the dried network was ~ 2.5 μL.

The results shown in Figure 5 indicate that the total saturation generally evolves linearly with time for a liquid flow rate between 0.5 and 0.01μL/min. From Figure 5, we can also estimate that the actual flow rates were: 0.6μL/min and 0.012μL/min for curves $a$ and $b$, respectively. The difference between the actual and programmed flow rates is attributed to the effective elasticity of the network delivery system (PDMS elasticity), uncertainty associated with the volume of individual channels, and inaccuracies in the pump delivery system.

### 4.2 Dynamics and distribution of the local saturation, $s$

The local saturation, $s$, provides a detailed description of the two-phase dynamics in the pore network during the filling process. Local saturation profiles measured using the method presented in Figure 4 (a-2) and (c) are similar to those presented in Figure 6. The curve corresponding to the first image $n=1$ (red) in the stack of patterns plotted in Figure 6 (a) and (b) denotes time ($t=0$) of water invasion, and the pink curve corresponds to the breakthrough moment, $P$. The index, $n$, of the image in the sequence is a measure of time since the image acquisition time step is a constant.

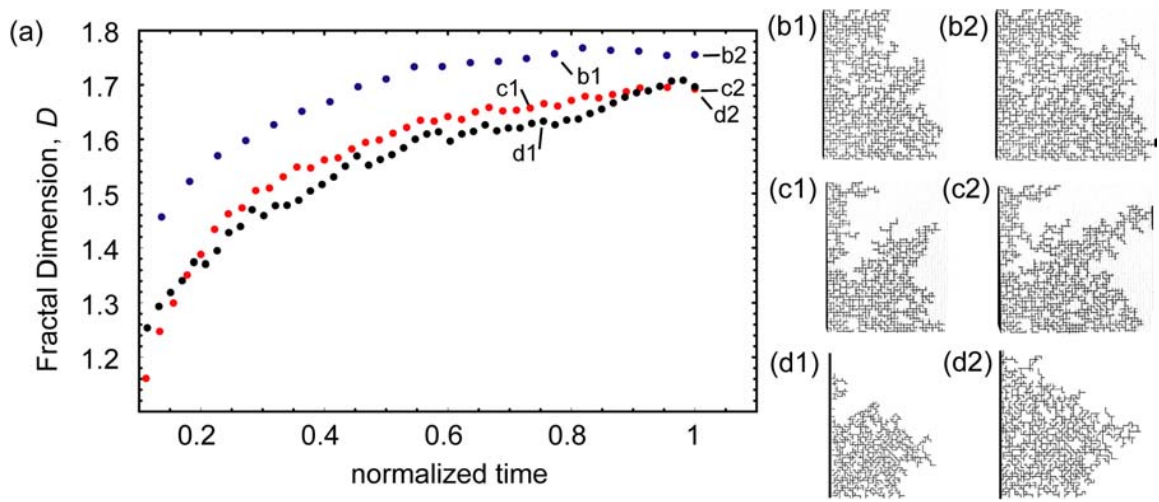

**Figure 8:** Evolution of fractal dimension of invading patterns with respect to time. Blue, red, and black points represent the following flow conditions: 0.5 μL/min, 0.01 μL/min, and quasi-static numerical simulation. Corresponding network images are shown at right, with indices indicating associated data points on (a).

Figure 6 shows the local saturation, $s$, versus the column index, $x$, for two different flow rates: Q=0.01 μL/min (a) and Q=0.5 μL/min (b). The local saturation, $s$, never reaches the maximum value of $s = 1$. This is an inherent property of hydrophobic networks; hydrophilic networks in our experiments exhibited much higher local saturation values $s \sim 1$. We also observed that the filled channels were not homogeneously distributed across the pore network. Saturated channels were concentrated at the inlet. This effect was more pronounced with decreasing flow rates. The flow patterns shown in Figure 6 (a) and (b) can be associated with the GDL, where the inlet lies adjacent to a saturated catalyst layer, and water percolates through the GDL until it reaches the gas channel interface at the breakthrough point. Furthermore, the increase in flow rate dramatically affects the density of the flow pattern. The amount of trapped air decreases significantly with increasing flow rate as shown by comparing Figure 6 (c) and (d).

The saturation profiles for both flow rate cases at breakthrough are plotted in Figure 7. The profiles show some similar trends owing to the geometric similarity of the networks. The increased flow rate, however, results in an increased saturation in the higher flow rate case that is approximately consistent throughout the network. It is also noteworthy that very low saturation points (s < 0.1) in the lower flow rate case are the most influenced by the increase in flow rate as the local velocities are higher (and local capillary numbers are higher), at these constricted parts of the network.

**4.3 Flow pattern fractal dimension**

We measured the fractal dimensions, $D$, of the flow patterns and compared the pattern structures and fractal dimensions with data presented in (13, 35) to characterize the measured flow patterns. The method presented in Figure 4 (a-3) and (d) was employed to calculate fractal dimensions for a series of flow patterns. Figure 8 (b) presents $D$ with respect to time for two different flow rates: 0.5 μL/min and 0.01 μL/min. The numerical results obtained from the quasi-static pore network model (32) are also presented in Figure 8. The normalized time is defined as the index of the image in the stack normalized with the index of the image corresponding to breakthrough.

Generally, we observe that *i*) during the initial invasion period, the flow pattern structures evolved similarly for different flow rates, *ii*) the fractal dimension reaches its steady state, $D_s$, before breakthrough at the outlet, and *iii*) different flow rates, $Q$, provided different $D_s$ values (Figure 8 (a)).

We found that high flow rates (2 - 10 μL/min) yielded dense clusters that exhibit a smooth invasion front, while low flow rates (> 1μL/min) produce liquid clusters with structures similar to those observed by Lenormand et al. (1988). Since our objective is to model the flow conditions similar to the GDL during PEMFC operation, we limit our presentation to low $Ca$ flows.

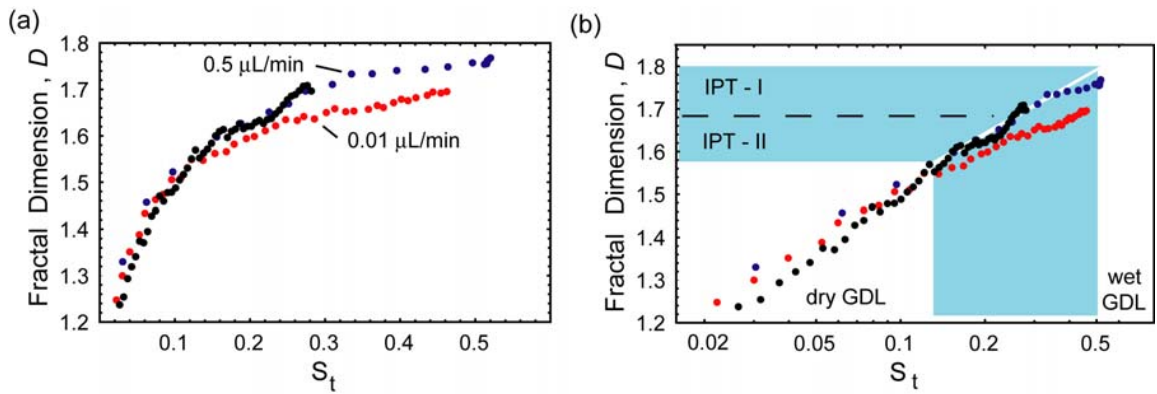

**Figure 9:** Fractal dimension versus total saturation diagrams: linear (a) and semi-log (b). Blue, red, and black points represent the following flow conditions: 0.5 μL/min, 0.01 μL/min, and quasi-static numerical simulation. The blue zones depict the interval of parameters of interest for GDL applications; for abbreviations see the text.

The typical flow clusters for low flow rates are presented in Figure 8. Depending on the flow rate, these clusters either occupy most of the network volume (b1-b2) (Type I, invasion with high flow rates) or aggregate into spatial structures (c1-c2) (Type II, invasion with low flow rates). In all cases the liquid clusters trap the secondary (air) phase.

Dense clusters (Type I) fully spread over the accessible network, and their fractal dimension approaches $D \sim 1.8$. Thus, they may be classified as IPT fractals in accordance with the fractal invasion classification provided by (7, 13, 35). Type II clusters which correspond to the much lower flow rates have noticeably lower fractal dimensions, $D \sim 1.6\text{-}1.7$ (Figure 8 (c1-c2)). The theoretical flow patterns shown in Figure 8 (d1-d2) have a fractal dimension similar to those of Type II, so it is reasonable to take this low value $\sim 1.6$ as a limiting value for our IPT invasion. It is noteworthy that this value is significantly lower than was reported earlier (7, 13, 35).

## 5. Discussion
### 5.1 Comparing local saturation, $s$, with 3D simulations from literature

In this section, we compare our local saturation profiles during network invasion and at the breakthrough moment to profiles obtained numerically by (21) using pore network simulations. These authors calculated the 3D saturation profiles for a realistic fibrous GDL structure under different flow rates and initial conditions. Our experimental conditions correspond to the invasion case considered in (21) with "zero surface coverage," meaning the GDL was initially dry prior to invasion. Our results presented in Figures 6 and 7 qualitatively agree with those presented in (Sinha and Wang 2007). In both studies, the local saturation exhibits: *i)* similar unsteady dynamics and *ii)* similar inlet plateaus that are pronounced at high flow rates. With respect to differences, the characteristic capillary numbers are significantly different for (21) and our experiments. For example, in our case the plateau-like profiles (Figure 7) correspond to $Ca \sim 10^{-7}\text{-}10^{-8}$ whereas those of (21) are for $Ca \sim 10^{-3}\text{-}10^{-4}$. The large difference in $Ca$ numbers for the qualitatively similar saturation profiles is unclear, but it may be attributed to the difference in pore space topology. Differences in the capillary number range for these two approaches may also be due to dimensionality; in order to facilitate imaging the microfluidic network is planar whereas the virtual network employed by (21) includes transport in the depth direction. It is also important to note that these liquid water breakthrough patterns do not closely resemble the theoretical frameworks proposed by Nam and Kaviany (2003) or by Litster et al. (2006), which were discussed in section 1.2.

### 5.2 Range of fractal dimensions for IPT invasion

It is useful to analyze the data presented in Figure 8 (a) and (b) in light of the previously established classification of fractal patterns in porous media (13, 14, 18). According to these previous studies, two parameters govern pattern classification: viscosity ratio ($M \sim 50$) and capillary number ($Ca \sim 10^{-7}$-$10^{-8}$). For the cases considered here, the viscosity ratio is set by the choice of fluids, while the capillary number is a parameter that can change with flow rate. Generally, for $M \sim 50$ the flow patterns fall either in the regime exhibiting a smooth front invasion corresponding to high flow rates (high $Ca$) or in the regime exhibiting characteristic *capillary fingering* clusters corresponding to low flow rates (low $Ca$) (13, 21). A detailed mapping of invasion regimes was performed experimentally and numerically by (13). Lenormand et al. (1988) established that only IPT invasion corresponds to the capillary fingering domain.

The same group also measured a fractal dimension of $\sim 1.8$ for their IPT patterns (35), where $M$ was significantly less than 1. The similarity of IPT patters with varying $M$ was later recognized (13), and it was established that IPT patterns could be identified with a fractal dimension of $\sim 1.8$. In this work, we found that IPT invasion percolation patterns can be characterized by a range of fractal dimensions (1.6-1.8) depending on the flow rate (Figure 8), which is in contrast to only a single universal value. Furthermore, our lower fractal dimension limit of 1.6 is significantly lower than the 1.8 value reported earlier for IPT patterns (7, 13, 35). It is also interesting to note that in contrast to the case studied in (35), where $M \sim 10^{-3}$, and $S_t$ and $D$ increased as $Ca$ decreased, we observe that $S_t$ and $D$ are directly proportional to $Ca$.

In summary, we suggest that while Lenormand et al. data can provide much insight, it should be applied with caution when analyzing flow in GDL-like porous networks which exhibit highly hydrophobic interior surfaces with pronounced pinning effects. In the experiments performed by Lenormand et al., hydrophobicity was established with the use of non-wettable fluids that would not experience contact angle hysteresis and pinning effects that differ from those expected in GDL materials.

### 5.3 Saturation-fractality diagram for hydrophobic pore networks

By combining the total saturation, $S_t$, and the fractal dimension, $D$, the data for various flow rates can be compiled in one plot, as shown in Figure 9. Figure 9 demonstrates how the flow pattern evolves depending on the flow rate employed. After the initial injection, the fractality of the flow pattern increases in the first period of invasion (Figure 8 (a)). During invasion, the fractal dimension reaches a steady state value, as shown in Figure 8 (a) and Figure 9 (a), even though the liquid pattern continues to grow. Furthermore, the steady state fractal dimension, $D_s$, increases with increasing flow rate, with a minimum value exhibited by the quasi-static simulation (Figure 8 and 9, black points).

If we shadow the interval of the fractal dimension between $D_{IPT-I}$ and $D_{IPT-II}$, and reflect this area on the coordinate, $S_t$, through the curves plotted on Figure 9 (b), we obtain an interval of saturation and fractal dimension for all possible IPT patterns observed in our pore networks. We propose to utilize this map to correlate the fractal dimension of a GDL flow pattern to the total saturation of the material. Furthermore, we can correlate the total saturation with either a relatively dry or wet regime.

In the study presented here, the initial conditions mimic a saturated catalyst layer/GDL interface. For these conditions we thus expect that flow patterns near the catalyst layer/GDL interface will likely fall within the IPT of type 1 (IPT-I) regime. Away from this wet sub-layer, we suggest that the GDL interior should be designed to achieve dry operation (IPT-II).

### 6. Conclusion

Microfluidic pore networks designed to experimentally model PEMFC gas diffusion layers were fabricated to allow the direct optical visualization of liquid water transport. The pore networks were fabricated using a rapid-prototyping method involving soft lithography and PDMS elastomers. The use of microfluidic pore networks provides a promising alternative for studying liquid water transport in GDLs, as microfluidic pore networks can be designed and built rapidly and inexpensively. Moreover, the pore and throat shape and size distributions in such networks can easily be varied and used in conjunction with a theoretical model to provide iterative feedback.

When employing low flow rates for pore networks with a saturated inlet, characteristics parameters corresponding to realistic fuel cell operating conditions were achieved and repeatable flow patterns were obtained. It was found that the local saturation decreases toward the outlet direction with a dependence on flow rate. The flow patterns are characterized by the fractal dimension, which depends on the injection flow rate and reaches a steady value during the filling process. It was also found that the fractal dimensions of IPT invasion patterns range from 1.6 to 1.8, in contrast to previously established literature. The saturation and fractal dimension of the invading patterns were correlated and the possible operation conditions for a GDL mapped on a single diagram, providing guidance for the design of GDL materials for drier bulk conditions.

**References**

1. K. S. Udell, *International Journal of Heat and Mass Transfer*, **28**, 485 (1985).
2. J. T. Gostick, M. W. Fowler, M. A. Ioannidis, M. D. Pritzker, Y. M. Volfkovich and A. Sakars, *Journal of Power Sources*, **156**, 375 (2006).
3. E. C. Kumbur, K. V. Sharp and M. M. Mench, *Journal of the Electrochemical Society*, **154**, B1305 (2007).
4. I. Fatt, *Petroleum Transaction, AIME*, **207**, 144 (1956).
5. M. J. Blunt, *Current Opinion in Colloid & Interface Science*, **6**, 197 (2001).
6. M. Blunt, M. J. King and H. Scher, *Physical Review A*, **46**, 7680 (1992).
7. R. Lenormand and C. Zarcone, *Transport in Porous Media*, **4**, 599 (1989).
8. J. Bonnet and R. Lenormand, *Revue De L Institut Francais Du Petrole*, **32**, 477 (1977).
9. J. D. Chen and D. Wilkinson, *Physical Review Letters*, **55**, 1892 (1985).
10. M. Ferer, C. Ji, G. S. Bromhal, J. Cook, G. Ahmadi and D. H. Smith, *Physical Review E*, **70** (2004).
11. J. P. Stokes, D. A. Weitz, J. P. Gollub, A. Dougherty, M. O. Robbins, P. M. Chaikin and H. M. Lindsay, *Physical Review Letters*, **57**, 1718 (1986).
12. D. G. Avraam, G. B. Kolonis, T. C. Roumeliotis, G. N. Constantinides and A. C. Payatakes, *Transport in Porous Media*, **16**, 75 (1994).
13. R. Lenormand, E. Touboul and C. Zarcone, *Journal of Fluid Mechanics*, **189**, 165 (1988).
14. Y. C. Yortsos, B. Xu and D. Salin, *Physical Review Letters*, **79**, 4581 (1997).
15. T. C. Halsey, *Physics Today*, **53**, 36 (2000).
16. D. Wilkinson and J. F. Willemsen, *Journal of Physics a-Mathematical and General*, **16**, 3365 (1983).
17. M. Sahimi, *Reviews of Modern Physics*, **65**, 1393 (1993).
18. J. Feder, *Fractals*, p. 281, Plemun Press, New York (1988).
19. B. Markicevic, A. Bazylak and N. Djilali, *Journal of Power Sources*, **171**, 706 (2007).
20. J. T. Gostick, M. A. Ioannidis, M. W. Fowler and M. D. Pritzker, *Journal of Power Sources*, **173**, 277 (2007).
21. P. Sinha, K., and C.-Y. Wang, *Electrochimica Acta* (2007).
22. O. Chapuis, M. Prat, M. Quintard, E. Chane-Kane, O. Guillot and N. Mayer, *Journal of Power Sources*, **178**, 258 (2008).
23. K. W. Feindel, L. P. A. LaRocque, D. Starke, S. H. Bergens and R. E. Wasylishen, *Journal of the American Chemical Society*, **126**, 11436 (2004).
24. K. Teranishi, S. Tsushima and S. Hirai, *Journal of the Electrochemical Society*, **153**, A664 (2006).
25. N. Pekula, K. Heller, P. A. Chuang, A. Turhan, M. M. Mench, J. S. Brenizer and K. Unlu, *Nuclear Instruments & Methods in Physics Research Section a-Accelerators Spectrometers Detectors and Associated Equipment*, **542**, 134 (2005).
26. I. Manke, C. Hartnig, M. Grunerbel, W. Lehnert, N. Kardjilov, A. Haibel, A. Hilger, J. Banhart and H. Riesemeier, *Applied Physics Letters*, **90** (2007).
27. J. H. Nam and M. Kaviany, *International Journal of Heat and Mass Transfer*, **46**, 4595 (2003).
28. S. Litster, D. Sinton and N. Djilali, *Journal of Power Sources*, **154**, 95 (2006).
29. V. Berejnov, N. Djilali and D. Sinton, *Lab on a Chip*, **8**, 689 (2008).
30. Y. N. Xia and G. M. Whitesides, *Annual Review of Materials Science*, **28**, 153 (1998).
31. J. C. McDonald, D. C. Duffy, J. R. Anderson, D. T. Chiu, H. K. Wu, O. J. A. Schueller and G. M. Whitesides, *Electrophoresis*, **21**, 27 (2000).
32. A. Bazylak, V. Berejnov, B. Markicevic, D. Sinton and N. Djilali, *Electrochimica Acta*, **53**, 7630 (2008).
33. B. Markicevic and N. Djilali, *Physics of Fluids*, **18** (2006).
34. D. C. Duffy, J. C. McDonald, O. J. A. Schueller and G. M. Whitesides, *Analytical Chemistry*, **70**, 4974 (1998).
35. R. Lenormand and C. Zarcone, *Physical Review Letters*, **54**, 2226 (1985).